\documentclass[floats,floatfix,showpacs,preprint,prd,superscriptaddress,onecolumn,nofootinbib]{revtex4}
\usepackage{graphicx, epsfig, amssymb} %include figure files
\usepackage{amsmath, amsfonts}
\usepackage{bm} %include bold math: \bm{} creates bold letters in math mode
\usepackage{color}
\usepackage[usenames]{xcolor}

\newcommand{\non}{\nonumber}
\newcommand{\be}{\begin{equation}}
\newcommand{\ee}{\end{equation}}                  
\newcommand{\bea}{\begin{eqnarray}}
\newcommand{\eea}{\end{eqnarray}}

\newcommand{\ord}[1]{\mathcal{O}\left(#1\right)}
\begin{document}

%%%%%%%
 
\title{Slowly rotating neutron stars in the nonminimal derivative coupling sector of Horndeski gravity.}

%%%%%%%

\author{Adolfo Cisterna}
\email{adolfo.cisterna@uach.cl}
\affiliation{Instituto de Ciencias Fis\'icas y Matem\'aticas,\\ Universidad Austral de Chile,\\ Valdivia, Chile}
\affiliation{Departamento de F\'isica, Universidad de Concepci\'on, Casilla, 160-C,\\ Concepci\'on, Chile}
\author{T\'erence Delsate}
\email{terence.delsate@umons.ac.be}
%\affiliation{Theoretical and Mathematical Physics Dept.\\ University of Mons - UMONS\\ 20, Place du Parc 
\author{Ludovic Ducobu}
\email{ludovic.ducobu@student.umons.ac.be}
\affiliation{Theoretical and Mathematical Physics Department, University of Mons \\ 20, Place du Parc - 7000 Mons, Belgium}
\author{Massimiliano Rinaldi}
\email{massimiliano.rinaldi@unitn.it}
\affiliation{Dipartimento di Fisica, Universit\`a di Trento,\\  Via Sommarive 14, 38123 Povo (TN), Italy}
\affiliation{TIFPA - INFN,\\  Via Sommarive 14, 38123 Povo (TN), Italy}

\begin{abstract}

%%%%%%%%
\noindent 
This work is devoted to the construction of slowly rotating neutron stars in the framework of the nonminimal derivative coupling sector of Horndeski theory. We match the large radius expansion of spherically symmetric solutions with cosmological solutions and we find that  the most viable model has only one free parameter. Then, by using several tabulated and realistic equations of state, we establish numerically the upper bound for this parameter in order to construct neutron stars in the slow rotation approximation with the maximal mass observed today.  We finally study the surface redshift and the inertia of these objects and compare them with known data.
\end{abstract}
%%%%%%

\maketitle

%%%%%%

%%%%%%%%%%%%%%%%%%%%%%%%%%%%%%%%
\section{Introduction}
%%%%%%%%%%%%%%%%%%%%%%%%%%%%%%%%

\noindent Nowadays, it is a well-accepted point of view that the validity of General Relativity (GR) is constrained by an ultraviolet (UV) cut-off of the order of the Planck scale.  On the other hand, at low energy regimes GR becomes increasingly accurate not only in the Solar System but also in the strong-field regime, as witnessed by the recent discovery of gravitational waves generated by the merger of two black holes \cite{GW}.

Nevertheless, despite the enormous success of GR in describing the nature of the gravitational interaction \cite{Damour, Turyshev:2008ur},  based on the simplicity of its principles and the successful experimental tests, there are still a number of phenomena for which the theory is not able to give a meaningful explanation. These phenomena are not only related to the UV regime but also to processes occurring at infrared (IR) energy scales. This is the main reason why theorists have tried different roads towards the construction of an extension of GR that is not in contrast with experimental data. 

Historically, long before their experimental discovery in the form of the Higgs field \cite{atlas}, scalar fields have played a fundamental role in the theoretical construction of new theories. Considering the UV scale,  string theory contains several scalar fields as  essential elements of its scaffolding. In particular, the dilaton field appears as an irreducible representation of the first excited states \cite{Polchinski:1998rq}-\cite{Tong:2009np}. Scalar fields also arise naturally in others higher dimensional theories like in Kaluza-Klein theory, where gravity and electrodynamics can be formulated as different manifestations of the gravitational field in a five-dimensional Universe. As a consequence of the dimensional reduction a scalar field appears \cite{Fon9}. Scalar fields finally appears frequently in supersymmetry and supergravity theories as auxiliary fields ensuring off-shell realisation of supersymmetry \cite{Freedman:2012zz}.
%Say something related to solitons and particle physics.

At IR energy scales the situation is similar. Since the very beginning of GR, the cosmological constant have been one of the most debated part of the theory. The discovery of the cosmic acceleration seemed to validate the existence of a fundamental cosmological constant  \cite{expansion}. However, its extremely small value, compared to the predictions from quantum field theory seems to indicate that GR needs to be modified also at IR scales \cite{Weinberg:1988cp}.

One of the simplest ways to modify gravity is the inclusion of new degrees of freedom. When these are in the form of one or more scalar fields, we have the so-called scalar-tensor theories of gravity. The first example was studied in the pioneering work of Brans and Dicke in 1961 \cite{BD}. The possibility of constructing compact objects in the context of Brans-Dicke theory was explored in many works, beginning from the exotic ``boson stars'' \cite{bosestar} up to  neutron stars with minimal or non-minimal coupling of the scalar field to matter, see e.g. \cite{nsscalar}. Finally, the most general scalar-tensor theory constructed with a single scalar field, in four spacetime dimensions, and with at most second-order equations of motion was discovered by Horndeski in the early seventies \cite{Horndeski}, but this work remained in the shadow for decades, until it was rediscovered in terms of the so-called ``Galileon'' (see below). 

In general, to be in agreement with observational and experimental data, any new degree of freedom is expected to modify gravity at large cosmological scales, but, at the same time, it is strongly suppressed at scales of the order of the Solar System. This calls for some screening mechanism able to ``hide'' the scalar field at short distances. One of the first modifications of gravity with a proper screening mechanism for the scalar field is the so-called Dvali-Gabadadze-Porrati (DGP) model \cite{DGP}. This higher dimensional model is based on the existence of a 3-brane surface embedded on a 5D Minkowski spacetime. In contrast to Kaluza-Klein scenarios, the extra dimension has an infinite size. Together with the usual GR action in 5 dimensions,  a scalar curvature term on the brane, induced by matter fields  living on the brane, is included. The outcome of this construction is  that, from a 4-dimensional point of view, gravity is mediated by a massive graviton and one scalar degree of freedom. The standard gravitational potential is recovered at small distances, while  a fully 5-dimensional potential dominates when the scales are larger than a specific crossover limit. This model was extensively analyzed due to its interesting cosmological solutions \cite{Deffayet:2000uy, Lue:2005ya, Deffayet:2001pu}. In particular, one solution branch shows a self-accelerating behavior without any cosmological constant term. In the decoupling limit, the theory exhibits an effective scalar field theory, with equations of motion of second order, which turns out to be invariant under Galilean transformations\footnote{The theory is invariant under transformations of the type $\phi\rightarrow \phi+\phi_{0}+b_{\mu}x^{\mu}$, where $\phi_{0}$ and the vector $b_{\mu}$ are constants.}, and which naturally includes  the screening Vainhstein mechanism \cite{Babichev:2013usa}. Soon after these developments, the decoupling limit of the DGP model was generalized into  the  Galileon theory \cite{Nicolis:2008in}.  Following the standard minimal coupling procedure, the covariantized version of Galileon gravity was constructed in \cite{galileon1}, where it was shown that the resulting theory possesses equations of motion of third order.
Nevertheless in the same work, it was shown that, including proper nonminimal couplings between the scalar field and curvature terms,  the second order character of the theory can be recovered \footnote{The D-dimensional version of the theory was obtained in \cite{Deffayet:2009mn}.}. Later it was shown that this theory is equivalent to Horndeski theory and  that its Lagrangian can be cast in a very simple form \cite{Kobayashi:2011nu}. In this framework, it is convenient to partition the Lagrangian of Horndeski gravity into sub-sectors according to
\begin{equation}\label{Haction}
S=\sum_{i=2}^{5}\int d^{4}x\sqrt{-g}{\cal L}_i\ ,
\end{equation}
where
\bea
{\cal L}_2&=&G_2\ ,\\\non
{\cal L}_3&=&-G_{3}\square\phi\ ,\\\non
{\cal L}_4&=&G_{4}R+G_{4{X}}\left[(\square\phi)^2-(\nabla_{\mu}\nabla_{\nu}\phi)^2\right]\ ,\\\non
{\cal L}_5&=&G_{5}G_{\mu\nu}\nabla^{\mu}\nabla^{\nu}\phi-\frac{G_{5{X}}}{6}\left[(\square\phi)^3+2(\nabla_{\mu}\nabla_{\nu}\phi)^3\
-3(\nabla_{\mu}\nabla_{\nu}\phi)^2\square\phi\right]\ .
\eea
Here, $G_i$ are arbitrary functions of the scalar field and of its canonical kinetic term $X\equiv -\nabla_{\mu}\phi\nabla^{\mu}\phi$ while $G_{iX}$ denote their derivatives with respect to $X$.

In cosmology, Horndeski gravity became very popular for its self-tuning property that allows to circumvent Weinberg's theorem on the cosmological constant \cite{fabfour}. Shortly after, it was discovered that the non-minimal kinetic coupling sector ${\cal L}_5$ (called ``John'' in the ``Fab Four'' terminology of \cite{fabfour}) leads to an accelerated expansion provided $G_{5}$ is constant and $G_{5X}=0$, see e.g.  \cite{Germani:2010gm, Amendola:1993uh, Sushkov:2009hk, Myrzakulov:2015ysa, Gumjudpai:2015vio, namur} and references therein. 
A lot of work was also done in perturbation theory, with the goal of finding potentially observable deviations from GR in large-scale structures and the conditions on the parameter space that avoid too large gravitational instabilities \cite{pert}.

As mentioned above, any modification of GR must be consistent with constraints at the Solar System level, which are very stringent. In order to verify such compatibility, it is important to study also spherically symmetric solutions of the theory, starting with black holes. Initially, such solutions appeared to be severely  constrained by the existence of a non-hair theorem \cite{Hui:2012qt}. However,  static black hole solutions with asymptotically anti-de Sitter behavior were found in the following sub-sector of the Horndeski action given by 
\begin{equation}
S=\int d^{4}x\sqrt{-g}\left[\kappa(R-2\Lambda)-\frac{1}{2}\left(\alpha g_{\mu\nu}-\eta G_{\mu\nu} \right)\nabla^{\mu}\phi\nabla^{\nu}\phi \right], \label{model}
\end{equation}
where $\kappa=(16\pi G)^{-1}$, $\alpha$ and $\eta$ are two parameters controlling the strength of the minimal and nonminimal kinetic couplings  \cite{rinaldi, Kolyvaris:2011fk, charmousis1, Kobayashi:2014eva, adolfo2, Minam1, adolfo3, minas}. One important feature of this model is that the shift symmetry $\phi\rightarrow\phi +\phi_0$ implies that the equation of motion for the scalar field can be written as the current conservation law $\nabla_{\mu}J^{\mu}=0$.

The first static exact black hole solution with $\Lambda=0$ and $\alpha=1$ was found in \cite{rinaldi}. This solution has one regular horizon and $\phi'^{2}<0$ outside it (from now on,  the prime indicates a derivative with respect to the radial coordinate). However, this does not implies any thermodynamical instability (besides a standard Hawking-Page transition) because the physical scalar degree of freedom is $\phi'^{2}$ and not $\phi'$. A more general solution that admits a scalar field that is real everywhere is obtained when $\Lambda<0$, as shown in \cite{adolfo2}. In particular,  for any combination of parameters such that $\alpha+\Lambda\eta<0$ the scalar field turns out to be real. In addition, the thermodynamical analysis revealed the existence of a Hawking-Page transition between a thermal soliton and large hairy black hole configuration \footnote{This family of solutions is not continuously connected with the (anti)-de Sitter maximally symmetric spacetime. However, there is a unique regular solution when the mass parameter  vanishes.}. For further studies of the thermodynamics of these black holes, see \cite{Feng:2015oea, Peng:2015yjx}.

Along the same lines, a more general family of solutions was presented in \cite{charmousis1} where the scalar field is time-dependent according to $\phi(t,r)=Qt+F(r)$, for some function $F$ and a constant $Q$. The scalar degree of freedom no longer shares the same symmetries than its tensorial partner  but it maintains a static contribution to the equations of motion. 
%In this solution, $J^{r}=0$ is required to satisfy the $tr$ component of the Einsteins equations. 
%Solutions \cite{rinaldi} and \cite{adolfo2} are recovered in the appropriated limits. 

A natural step forward in the investigation of this sector of Horndeski gravity is to study compact objects, in particular neutron stars.  These astrophysical objects  typically have mass and radius of the order of $1.3 \div 2$  solar masses and $8 \div15$ kilometres respectively so they are extremely dense. These features make them excellent candidates to probe the strong field regime and, hopefully, to find observable deviations from standard GR \cite{Berti, Schutz}. 

As far as we know, the first attempt to build this kind of configuration for the system (\ref{model}) was proposed in \cite{Cisterna:2015yla}. There,  neutron stars and white dwarfs were shown to exist and constraints on the only free parameter of the model (namely the product $Q^{2}\eta$) were found. One of the most attractive and surprising  features of these solutions is that the metric outside the surface of the star is identical to the Schwarzschild metric therefore there are  no conflicts with Solar System tests.

The results presented in \cite{Cisterna:2015yla} were limited to static configuration. In the present work, we investigate the structure of a rotating star for the theory (\ref{model}) using realistic equations of state. The plan of the paper is the following: in Section \ref{sec2} we review the spherically symmetric solutions constructed in \cite{charmousis1} that were used to construct static configurations. Section \ref{sec3} is devoted to the cosmological solutions of the theory \eqref{model}.  Here we will focus on  the matching of the cosmological constants obtained from spherically symmetric solutions and from cosmological ones to show that the two are compatible and that the approximations that we will use (namely $\Lambda=0$) are well-justified. In Section \ref{sec4} we study the equations for the slowly rotating neutron stars. In Section \ref{sec5} we present the results of numerical computations and we compare them to some astrophysical data. We conclude in Section \ref{sec6} with some remarks.

%%%%%%%%%%%%%%%%%%%%%%%%%%%%%%%%
%%%%%%%%%%%%%%%%%%%%%%%%%%%%%%%%

%\noindent \emph{Notations} 

%\noindent Before detailing our results, we introduce some notations. For astrophysical applications, we will use the metric
%\be
%ds^2 = -b(r) dt^2 + \frac{dr^2}{f(r)} + r^2d\Omega^2,
%\label{ds2astro}
%\ee
%whereas for cosmological solutions we will use the metric
%\be
%ds^2 = -dt^2 + a(t)^2 (dr^2 + r^2 d\Omega^2).
%\label{ds2cosmo}
%\ee
%In both cases, we will eventually assume that the  scalar field is given by
%\be\label{phians}
%\phi(t,r) = Q t + \phi_0(r).
%\ee
%where $\phi_0(r) = \phi_0$ is a constant in the case of cosmological solutions. The parameters $\Lambda$, $\kappa$, and $Q$ will be often called ``bare'' as their physical value can change in a specific solution. In particular, and for the spherically symmetric solutions, the physical value of $Q$ is given by the value measured by a distant observers, i.e.  
%\be
%Q_p = \frac{Q}{\sqrt{b_\infty}},
%\ee
%where the constant $b_{\infty}$ is defined for large radii $r$ by  
%\be
%g_{tt} \approx b_\infty \left(-\frac{\Lambda_m}{3} r^2 + 1 - {2M\over r}\right),
%\ee
%and $\Lambda_m$ is the physical cosmological constant and $M$ is the physical mass of the compact object. For cosmological solutions however $Q\equiv Q_{p}$.

%%%%%%%%%%%%%%%%%%%%%%%%%%%%%%%%
\section{Vacuum spherically symmetric solutions}\label{sec2}
%%%%%%%%%%%%%%%%%%%%%%%%%%%%%%%%

\noindent In this section we review in detail the spherically symmetric solutions constructed in  \cite{charmousis1} and used in \cite{Cisterna:2015yla} for the modelling of static neutron stars.\\
The equations of motion coming from the action (\ref{model}) are given by
\begin{eqnarray}
&&G_{\mu\nu}+\Lambda g_{\mu\nu}+H_{\mu\nu}=0\,, \label{eqmetric}\\
&&\nabla_{\mu} J^\mu  =0\,, \label{eqphi}
\end{eqnarray}
where
\begin{eqnarray}
H_{\mu\nu}&=&-\frac{\alpha}{2\kappa}\Bigg[\nabla_{\mu}\phi\nabla_{\nu}\phi-\frac{1}{2}g_{\mu\nu}\nabla_{\lambda}\phi\nabla^{\lambda}\phi\Bigg]\nonumber-\frac{\eta}{2\kappa}\Bigg[\frac{1}{2}\nabla_{\mu}\phi\nabla_{\nu}\phi R-2\nabla_{\lambda}\phi\nabla_{(\mu}\phi R_{\nu)}^{\lambda}\nonumber\\
&-&\nabla^{\lambda}\phi\nabla^{\rho}\phi R_{\mu\lambda\nu\rho}-(\nabla_{\mu}\nabla^{\lambda}\phi)(\nabla_{\nu}\nabla_{\lambda}\phi)+\frac{1}{2}g_{\mu\nu}(\nabla^{\lambda}\nabla^{\rho}\phi)(\nabla_{\lambda}\nabla_{\rho}\phi)-\frac{1}{2}g_{\mu\nu}(\square\phi)^{2}\nonumber\\
&&+(\nabla_{\mu}\nabla_{\nu}\phi)\square\phi+\frac{1}{2}G_{\mu\nu}(\nabla\phi)^{2}+g_{\mu\nu}\nabla_{\lambda}\phi\nabla_{\rho}\phi R^{\lambda\rho}\Bigg] \,,\\
J^\mu &=&  \left(  \alpha g^{\mu\nu}-\eta G^{\mu\nu}\right)\nabla_{\nu}\phi\,.
\end{eqnarray}
The spherically symmetric metric is chosen as
\be
ds^2 = -b(r) dt^2 + \frac{dr^2}{f(r)} + r^2d\Omega^2.
\label{ds2astro}
\ee
The shift symmetry of the action allows static solutions with a linearly time-dependent scalar field of the form
\begin{equation}
\phi(t,r)=Qt+F(r)\ , \label{Qt}
\end{equation}
where $F(r)$ is an arbitrary function. This implies that even if the scalar field does not share the same symmetries of the spacetime background, the energy-momentum tensor does, avoiding in this way the non-hair theorem of \cite{Hui:2012qt} and allowing black holes configurations to have a scalar hair. In fact, the key point of \cite{Hui:2012qt} is that shift symmetric theories possess a scalar field equation given by a current conservation law (\ref{eqphi}). If we demand  that the norm of the scalar field current remains finite at the horizon, we find that $J^{r}=0$ at any point in the domain of outer communications. For shift symmetric theories like eq.\ (\ref{model}), where at least the Lagrangian contains term of second order in the scalar field gradient, the current can always be cast in the  form
\begin{equation}
J^{r}=\phi'\Theta(\phi',g,g',g'')\,,
\end{equation}
where $g, g'$ and $g''$ denotes the metric functions and their first and second derivatives. For asymptotically flat solutions, the function $\Theta$ tends to a constant value, forcing the scalar field to become trivial in order to satisfy the condition $J^{r}=0$, and naturally ruling out the existence of scalar hair. However for scalar fields like (\ref{Qt}), the Einstein equations include a non-trivial off-diagonal  component of the form $\mathcal{E}_{tr}(r)=0$. This equation holds if, and only if, $\Theta=0$, implying no conditions on the scalar field and satisfying, at the same time, the vanishing current norm condition on the horizon located at $r=r_{h}$, namely
\begin{equation}
|J|^{2}=-b(r_{h})\left(J^{t}\right)^{2}+\frac{\left(J^{r}\right)^2}{f(r_h)}=0\,.
\end{equation}
Then, configurations with non-trivial scalar field exist \footnote{Another interesting way to circumvent the non-hair conjecture of \cite{Hui:2012qt} is to considerer a sub sector of the shift invariance Horndeski Lagragian in which a linear coupling between the scalar field and the Gauss-Bonet density is considered \cite{Sotiriou:2013qea}.}.
Note that the condition $\Theta=0$ can be arbitrarily imposed considering (anti)-de Sitter asymptotic geometry for solutions where the scalar field depends exclusively on the radial coordinate \cite{rinaldi, adolfo2}\footnote{In contrast to the solutions found in \cite{rinaldi, adolfo2}, the scalar field  and its derivative (\ref{Qt}) not only are regular on the horizon but they are also analytic.}.
From the equation $\Theta=0$ we conveniently find one of the metric functions in terms of the other, namely
\begin{equation}
f(r)= \frac{ (\alpha r^2+\eta) b(r)}{ \eta (rb'(r)+b(r))}\,.
\end{equation}
Inserting  this relation into the $rr$-component of the Einstein equations, one finds 
\begin{equation}
F'(r) = \pm \frac{\sqrt{r}}{2b(r)(\alpha r^2+\eta)}\left[Q^2\eta (\alpha r^2+\eta) b'(r)-\kappa(\alpha+\eta\Lambda)(b(r)^2r^2)'\right]^{1/2}.
\end{equation}
Finally, inserting both relations above into the $tt$-component of the Einstein equations, yields a differential equation for $b(r)$. The authors of \cite{charmousis1} expressed this equation by introducing the implicit definition
\begin{equation} 
 b(r) = -\frac{\mu}{r} +\frac{2}{r}\int \frac{K(r)dr}{\alpha r^2+\eta}
\end{equation}
where $\mu$ is an integration constant. This allows to express implicitly the function $K(r)$ as the solution of the algebraic equation of third order given by
\begin{equation}\label{bigK}
\frac{Q^2\eta}{8} (\alpha r^2+\eta)^2 - \kappa\left[\eta+\left( \alpha -\frac{1}{2}(\alpha+\eta\Lambda) \right) r^2\right] K + C_0 K^{3/2} =0\,,
\end{equation}
where $C_0$ is another integration constant.
This algebraic equation is very difficult to solve for the most general case. However, there are some interesting and simple  solutions that can be obtained for specific choices of the parameters.
In the following, the quantities  $\Lambda$, $\kappa$, and $Q$ will be often called ``bare'' as their observable value can change a specific solutions. In particular, and for the spherically symmetric solutions, the physical value of $Q$ is given by the value measured by a distant observers, i.e.  
\be\label{qupi}
Q_p = \frac{Q}{\sqrt{b_\infty}}\,,
\ee
where the constant $b_{\infty}$ is defined, for large radii $r$, by  
\be
 g_{tt} \approx b_\infty \left(-\frac{\Lambda_m}{3} r^2 + 1 - {2M\over r}\right) \,.
\ee
Here, $\Lambda_m$ is the measured (or physical) cosmological constant, and $M$ is the physical mass of the compact object. For cosmological solutions however $Q\equiv Q_{p}$, see next section.\\
It can be shown that the lapse function $b$ can be written as
\be\label{asyb}
b(r) = 1 + \frac{\alpha  \left[\kappa  (\alpha -\eta  \Lambda )+\alpha  \eta Q_p^2\right]}{3 \eta  \kappa  (3 \alpha +\eta  \Lambda )}\,r^2 -\frac{2M}{r} \,,
\ee
and the shift function $f$ as 
\bea\label{asyf}
f(r) = \frac{\alpha  r^2}{3 \eta } + \frac{7 \alpha  \kappa +\eta  \kappa  \Lambda +\alpha  \eta  Q_p^2}{3 \alpha  \left(\kappa +\eta Q_p^2\right)-3 \eta  \kappa  \Lambda }+ \frac{2M \kappa  (3 \alpha +\eta  \Lambda )}{r \left[\kappa  (\alpha -\eta  \Lambda )+\alpha  \eta Q_p^2\right]}\,.
\eea
In order to avoid conical singularities, the constant term in $f(r)$ must be equal to one and this leads to the constraint on the bare cosmological constant given by
\be\label{bareL}
\Lambda = -\frac{\alpha}{\eta} \left(1-\frac{ Q_p^2\eta}{2\kappa}   \right)\,.
\ee
By substituting this result back into eqs.\ \eqref{asyb} and \eqref{asyf} we find
\be
f = b = \frac{\alpha}{3\eta} r^2 +1 -\frac{2M}{r} ,
\ee
which describes an asymptotic  Schwarzschild - (anti) de Sitter metric, with a mass parameter  $M$ and physical cosmological constant  
\be\label{Lm}
\Lambda_m = -\frac{\alpha}{\eta}.
\ee
Note also that
\be
 \frac{Q_p^2\eta}{2\kappa} = 1-\frac{\Lambda}{\Lambda_m}.
\ee
In the next section, we verify if the physical cosmological constant \eqref{Lm} is compatible with the one obtained by solving the equations of motion with a Robertson-Walker metric. In addition, since we are going to study rotating compact objects with $\Lambda=\alpha=0$, we want to be sure that these conditions are not incompatible with the cosmological solution.

By setting $\alpha=\Lambda=0$ eq.\ \eqref{bigK} can be easily solved yielding the vacuum solution corresponding to the so-called stealth configuration. This means that even if the scalar field has a non-trivial functional form, the tensor $H_{\mu\nu}$ in \eqref{eqmetric} vanishes identically. Then, the vacuum solution coincides with the Schwarzschild solution. However,  in the presence of matter fluids, this property no longer holds and this has important consequences for neutron star configurations. In particular, the metric outside the star is still exactly the same as the Schwarzschild one, avoiding conflicts with Solar System tests.

%%%%%%%%%%%%%%%%%%%%%%%%%%
\section{Cosmological solutions}\label{sec3}
%%%%%%%%%%%%%%%%%%%%%%%%%%

\noindent In this section we study the cosmological solutions obtained from the Lagrangian \eqref{model} implemented by the contribution of a perfect fluid. We choose the metric 
\be
ds^2 = -dt^2 + a(t)^2 (dr^2 + r^2 d\Omega^2)\,,
\label{ds2cosmo}
\ee
where $a(t)$ is the cosmological scale factor. In particular, we want to compare the resulting cosmological dynamics with the standard Cold Dark Matter model implemented with the cosmological constant (called in short $\Lambda$CDM). As mentioned in the previous section, black holes (as well compact objects) of this theory generically have an asymptotic  de Sitter or anti-de Sitter geometry. In the first case, we wish to compare the effective cosmological constant of these solutions to the one  that arises from cosmological solutions. We will see that these are the same in a dust-dominated Universe and do not coincide with the bare cosmological constant  $\Lambda$.

Using the metric \eqref{ds2cosmo}, we obtain from \eqref{model}  the Friedmann equations ($H=\dot a/a$)
\bea\label{Fried}
H^{2}&=&{2\rho+4\kappa\Lambda+\alpha\dot\phi^{2}\over 3\left(4\kappa-3\eta\dot\phi^{2}\right)}\,,\\\non\\\non
\dot H&=&{\rho\left[\eta\dot\phi^{2}(1+3\omega)-4\kappa(1+\omega)\right]\over (3\eta\dot\phi^{2}-4\kappa)(\eta\dot\phi^{2}-4\kappa)}-{2\dot\phi\left[ \eta H(3\eta\dot\phi^{2}-4\kappa)\ddot\phi+2\kappa(\alpha+\eta\Lambda)\dot\phi-\alpha\eta\dot\phi^{3}  \right]\over  (3\eta\dot\phi^{2}-4\kappa)(\eta\dot\phi^{2}-4\kappa)}\,,
\eea
where $\rho$ satisfies the usual equation for a perfect fluid
\bea\label{rho}
\dot \rho+3H\rho(1+\omega)=0\,.
\eea
Finally, there is the Klein-Gordon equation
\bea\label{KG}
\ddot\phi+3H\dot\phi\left(1+{2\eta\dot H\over \alpha+3\eta H^{2}} \right)=0\,.
\eea
It is easy to check that all the equations depend on $\dot\phi$ and thus are shift-invariant.

%%%%%%%%%%%%%%%%%%%%%%%%%%%%%%%%%%
\subsection{Inflationary solutions}
%%%%%%%%%%%%%%%%%%%%%%%%%%%%%%%%%%

\noindent Before studying the $\Lambda$CDM model, it is interesting to  look at the vacuum solutions $\rho=0$. This case was studied already  in  \cite{Sushkov:2009hk} and \cite{namur} for $\Lambda=0$ and $\alpha=1$. In the present case, it is convenient to find the effective equation of state parameter for the scalar field that reads
\bea
\omega_{\phi}&\equiv&-1-{2\dot H\over 3H^{2}}\\\non
&=&{ (-4\kappa+3\eta \phi^2) (-\eta\alpha^2 \phi^4-14\eta\alpha\kappa\Lambda\phi^2-2\alpha^2\kappa\phi^2+8\kappa^2\Lambda^2\eta+8\alpha\kappa^2\Lambda)  \over (\alpha \phi^2+4\kappa\Lambda)(3\alpha\eta^2\phi^4+6\eta^2\kappa\Lambda \phi^2-6\eta\alpha \kappa\phi^2+8\eta\kappa^2\Lambda+8\alpha\kappa^2)}\,.
\eea
In the high energy limit, which is identified with the regime where the kinetic term $\dot\phi^{2}$ is dominant, the equation of state approaches the value $-1$, which is required in order to have an inflationary phase. The advantage of this model, when compared to the usual single-field inflation, is that there is no need for an ``ad hoc'' potential for the scalar field: inflation naturally occurs whenever $\omega_{\phi}<-1/3$. Thus the inflationary phase exists just because $\eta\neq 0$.  However, it is easy to see that $\omega_{\phi}\rightarrow -1$ also when $\dot\phi$ vanishes  and this is  explained by the presence of the bare cosmological constant $\Lambda$, which takes over the dynamics at low energy  and in the absence of matter fluids (i.e. with $\rho=0$). The transition between the two accelerated phases is not physically viable because it seems that no reheating mechanism can be inserted without spoiling the full model.

If $\Lambda=0$ the equation of state parameter no longer depends on $\alpha$ and it varies from the value $-1$ at high energy to the value $+1$ at low energy. Therefore the model becomes the same as the one described in ref. \cite{namur}, where it is shown that there exists and inflationary phase followed by a graceful exit. However, in order to have a sufficiently long inflationary period, the initial value of $\dot\phi$ must be so large in Planckian units that quantum gravitational corrections cannot be ignored. In conclusion, the model \eqref{model} does not seem suitable for describing the inflationary Universe.

%%%%%%%%%%%%%%%%%%%%%%%%%%%%%%%%%%
\subsection{$\Lambda$CDM}
%%%%%%%%%%%%%%%%%%%%%%%%%%%%%%%%%%

\noindent We now study the equations of motion in a non-inflationary Universe and in the presence of a perfect fluid with energy density $\rho$, governed by eq.\ \eqref{rho}. Since we want to match these solutions with the large radius limit of the spherically symmetric solution, we set
\bea\label{lin}
\phi=Q t+\phi_{0}\,,
\eea
which reduces eq.\ (6) to
\bea
0&=&\alpha+\eta(2\dot H+3H^{2})\,.
\eea
By substituting  eqs.\ (3) and (4) and the ansatz \eqref{lin} we find the constraint
\bea
\Lambda={\alpha Q^{2}\over 2\kappa}+{\eta\omega\rho(t)-2\alpha \kappa\over 2\eta\kappa}\,,
\eea
which generically holds if either $\omega=0$ or $\rho(t)=0$ for all $t$. We choose the first option, which describes well our present Universe filled with cold dark matter and dark energy. Thus  the bare cosmological constant becomes
\bea
\Lambda=-{\alpha\over \eta}\left(1 - { Q^{2}\eta \over 2\kappa}\right)\,,
\eea
which is the same as \eqref{bareL}. Therefore, we have found the common value of $\Lambda$ such that spherically symmetric and cosmological solutions have the same physical cosmological constant. The important point is that the choice $\alpha=0$ necessarily implies $\Lambda=0$, which means that the neutron star models that we will present below are not in conflict with cosmological solutions.

By replacing this  expression  back into (3) and (4), implemented by the ansatz \eqref{lin}, we find
\bea
\dot H&=&-{\rho\over 4\kappa-3\eta Q^{2}}\,,\\\non\\
H^{2}&=&{2\rho\over 3(4\kappa-3\eta Q^{2})}-{\alpha\over 3\eta}\,.
\eea 
These equations coincide with the standard $\Lambda$CDM Friedmann equations
\bea
H^{2}&=&{\Lambda_{m}\over 3}+{\rho\over 6\tilde\kappa}\,\\\non\\
\dot H&=&-{\rho\over 4\tilde \kappa}\,,
\eea
where we identified the measured cosmological constant $\Lambda_{m}$ with
\bea\label{lambdames}
\Lambda_{m}=-{\alpha\over\eta}\,,
\eea
and 
\bea\label{kappaeff}
\tilde\kappa=\kappa-{3\eta Q^{2}\over 4}\,.
\eea
Note that eq.\ \eqref{lambdames}Ê implies that $\alpha$ and $\eta$ must have opposite signs. Note also that eq.\ \eqref{kappaeff} implies that
\bea
\eta< {4\kappa\over 3Q^{2}}\,,
\eea
which is the same upper limit found \cite{Cisterna:2015yla} from the requirement that the second derivative of the matter pressure inside a compact object  is negative. In turn, this is a necessary condition for the existence of compact objects. 

From these considerations, it appears that the parameter $\eta$ can be  arbitrarily large and negative. However, there is an argument to show that $\eta$ should be positive only. Consider the action \eqref{model} with $\alpha=0$ and suppose that the backreaction of the field is negligible so we can choose a de Sitter fixed background with $R_{\mu\nu}=\Lambda g_{\mu\nu}$. Suppose also that the field depends on time only, for simplicity. It follows that, on shell, the Lagrangian is $L\sim 2\kappa \Lambda+\eta \Lambda\dot\phi^{2}/2$ and the corresponding Hamiltonian is  ${\cal H}\sim \eta\Lambda\dot\phi^{2}/2-2\kappa\Lambda$. Therefore, a negative $\eta$ would give a negative definite Hamiltonian, which inevitably leads to instabilities upon quantization. Below we will consider negative $\eta$ in the numerical calculations but one should take this case with a grain of salt.

In summary, we have shown that cosmological solutions and spherically symmetric solutions have the same bare cosmological constant so there is no tension between the large scale geometry of the compact object and the cosmological evolution. This property holds only in a dust-filled Universe and for any value for the parameter $\alpha$, including $\alpha=0$, which implies $\Lambda=0$ in both neutron stars and current cosmological evolution \footnote{The case $\alpha=0$ does not necessarily means that the current physical cosmological constant  vanishes, which would be in contrast to observations. It simply means that the observed value of $\Lambda$ is due to other effects, such as contributions from the vacuum expectation value of quantum fields or extra degrees of freedom, such as quintessence.}.

%%%%%%%%%%%%%%%%%%%%%%%%%
\section{Slowly rotating neutron stars}\label{sec4}
%%%%%%%%%%%%%%%%%%%%%%%%%

\noindent We now focus on the main topic of this paper, namely stationary neutron stars with realistic equations of state. In particular, we study neutron stars and pulsars configurations within the slow rotation approximation (which is valid for about $80\%$ of the known pulsars) following the Hartle formalism to first order in rotation \cite{Hartle:1967he}. Very recently, some results along these lines were presented in  
\cite{Silva:2016smx} where a polytropic equation of state was considered. Our goal is to work instead with more realistic tabulated equations of state.

According to the prescriptions of \cite{Hartle:1967he}, we first generalise the metric \eqref{ds2astro} and the scalar field  according to 
\bea\label{rotmetric}
ds^2 &=& - b(r) dt^2 + \frac{dr^2}{f(r)}  + r^2 \left[d\theta^2+\sin^2 \theta \left( d\varphi  - \epsilon(\Omega_* - \omega(r) )dt)^2 \right) \right] ,\\\non\\
\Phi(t,r) &=& Q t + F(r) + \epsilon \phi_1(t,r),
\eea
where $\epsilon$ is a small ``book-keeping'' parameter and $\epsilon(\Omega_* - \omega(r) )$ is the coordinate angular velocity (to first order) of a fluid element as seen by a free falling observer from infinity.

The matter field is described by a perfect fluid, with the usual  stress energy tensor of the form
\be
T^{ab} = (\rho + P ) u^a u^b + P g^{ab},
\ee
where $P$ is the fluid pressure, and $\rho = \rho(P)$ is the energy density that depends on the pressure thought a (barotropic) equation of state. Finally, $u^a$ is the proper $4$-velocity of the fluid, given by
\be
 u_a=(-\sqrt{b},0,0,u_{\varphi}) \,,
\ee
where
\bea
u_{\varphi}={2\epsilon r^{2}\sqrt{b}(\omega-\Omega_{*})\sin^{2}\theta\over \Omega_{*}\left[b-(\omega-\Omega_{*})^{2}\epsilon^{2}r^{2}\sin^{2}\theta\right]}={2\sin^{2}\theta(\omega-\Omega_{*})r^{2}\epsilon\over\Omega_{*}\sqrt{b}}+\ord{\epsilon^2}
\eea
so that $u^2 = -1 + \ord{\epsilon^4}$.
The equations of motion (\ref{eqmetric}) and (\ref{eqphi}) are  modified by the matter source and now read
\begin{eqnarray}
G_{\mu\nu}+\Lambda g_{\mu\nu}+H_{\mu\nu}&=&( 2\kappa)^{-1}T_{\mu\nu}^{(m)}\,,\\ 
\nabla_{\mu} J^\mu &=&0\,,\\
\nabla_{\mu}T^{\mu\nu (m)}&=&0\,.
\end{eqnarray}
Using the metric \eqref{rotmetric}, and expanding to first order in $\epsilon$, we obtain 
\bea
rfb' &=&(1-f)b\,,\label{eqAB}\\\non\\\non
 \eta fb F'^{2}&=&(1-f)\eta Q^{2}+bPr^{2}\, ,\label{eqfp}\\\\\non
f'& =&{(f-1)(4\kappa b-3\eta Q^{2}f)+b(\rho+6fP+f\rho)r^{2}\over r\left[3\eta Q^{2}f - (Pr^{2}+4\kappa)b\right]}\,,\\\non\\
P'&=&-{b'(\rho+P)\over 2b}\,,
\eea
which are identical to the static case \cite{Cisterna:2015yla}. The effect of the slow rotation becomes apparent, even at the zeroth order in $\epsilon$, because the $(t\varphi)$-component of the Einstein equations yields a differential equation for $\omega$, which reads
\bea
\omega''(r) &=& \, K_{1} \omega' +K_{2} \omega\,,
\eea
where
\bea
K_{1} &=& { b'^{2}(P+\rho)r^{4}-2bb' (P-\rho)r^{3}-b^{2}(3P-\rho)r^{2} -16\kappa bb'r-4b(4\kappa b-\eta Q^{2}) \over br[(Pr^{2}+4\kappa)(b' r+ b)-\eta Q^2]}\,,\\\non\\
K_{2} &=& {8(b+rb')^2(\rho+P) \over b[(Pr^{2}+4\kappa)(b' r+ b)-\eta Q^2]}\,.
\eea
Finally, at the zeroth order in $\epsilon$, we also find that
\bea
 \phi_1(t,r) &=& 0\,,
\eea
which closes the system of equations.

Note that in vacuum ($\rho=P=0$), the equation for $\omega$ reduces exactly to its general relativistic counterpart  \cite{Cisterna:2015uya, Maselli:2015yva}. As a consequence, the vacuum solution for $\omega$ is the same as in GR, namely 
\be
\omega = \Omega_{*}\left( 1 - \frac{2I}{r^3} \right),
\label{eq:inertia}
\ee
where $I$ is the moment of inertia of the star \footnote{It is expected that this result no longer holds at higher order in the expansion.}.

In  \cite{Cisterna:2015yla} a polytrope equation of state was considered. Here, we use instead the following tabulated equations of state (EOS) described in \cite{EOS}: $BSk14$, $BSk19$, $BSk20$, $BSk21$, $SLy4$, and $EOSL$. These equations cover a wide range of nuclear parameters, although not all are consistent with astrophysical tests \cite{Fantinia:2013} within GR. 

Indeed, some of these EOSs do not reproduce the maximal neutron star observed mass (around $2M_\odot$), some do only marginally, while some other reach
the bound without problems. Some additional tests are discussed in \cite{Fantinia:2013}. The
reliability of the EOSs is summarized in TABLE I. 
\begin{table}
\begin{tabular}{lcc}
\hline
\hline
EOS 	&	Status within GR							&	reliability	\\
\hline
 BSk14	&	does not reach $2M_\odot$						&	$-$	\\
 BSk19	&	does not reach $2M_\odot$						&	$-$	\\
 BSk20	&	compatible with most observations					&	$+$	\\
 BSk21	&	compatible with most observations					&	$+$	\\
 SLy4	&	compatible with most observations, maximal mass close to $2M_\odot$	&	$\pm$\\
 eosL	& 	compatible for maximal mass						&	$+$\\
 \hline
\end{tabular}
\caption{We summarize the level of reliability of the EOS we consider in this paper. The BSK19-21 and SLy tests are taken from \cite{Fantinia:2013}, the BSK14 and eosL statuses are based on the ability to reproduce a 2 $M_\odot$ neutron star within GR. The column `reliability' is a qualitative level of reliability used in this paper: $+, \pm,$ and $-$ indicate respectively reliable, marginally reliable and not reliable.}
\end{table}

Even if some of the EOSs we use are disfavored within GR, we will include them in our
study for completeness.

It is important to point out that, as in the static case,  GR is not recovered in the limit $Q^2\eta \rightarrow 0$. This is related to the fact that $\eta$ is not a perturbative parameter, as already noted in \cite{rinaldi} for the black hole model. 

The parameter $\eta Q^{2}$ is constrained from cosmology, that gives the condition $Q^2\eta < 4 \kappa /3$ as seen in sec. \ref{sec3}. This constraint emerges again by expanding the equations of motion around the centre of the star. By imposing the usual requirements $b(0)=b_{0}>0$, $b'(0)=0$, $\phi(r=0)=0$, $P(0)=P_{0}>0$, $\rho(0)=\rho_{c}>0$ we find that
\bea
F'^{2}&=&{\eta Q^{2}(3P_{0}-2\rho_{0})-12\kappa b_{0}P_{0}\over 3\eta(3\eta Q^{2}-4\kappa b_{0})}\,r^{2}+\ord r^{4}\,,\\\non\\
P(r)&=&P_{0}+{b_{0}(3P_{0}+\rho_{0})(P_{0}+\rho_{0})\over 6(3\eta Q^{2}-4\kappa b_{0})}\,r^{2}+\ord r^{4}\,,
\eea
which are the same conditions found in the static case \cite{Cisterna:2015yla}. From these equations we desume that $Q^2\eta < 4 \kappa b_{0}/3$ since, in order to have a compact configuration, it is customary that $d^{2}P(r)/dr^{2}<0$ at $r=0$. In principle, when $\eta<0$ there is no upper bound on $Q^2 |\eta|$ but only a  lower bound that comes from the requirement that $F'^{2}(r)>0$. However, as discussed above, a negative value for $\eta$ seems incompatible with the quantum version of the theory. In any case, we will keep an open mind and consider also $\eta<0$ below. Without loss of generality, we will set $\eta =1,\ -1$  since all the equations depend only on the combination $Q^2\eta$ (except for the decoupled equation for  $F'$, see \eqref{eqfp}, where a redefinition of $\eta$ must be accompanied by suitable rescaling of the pressure, energy density, and $b(r)$.

%%%%%%%%%%%%%%%%%%%%%%%%%%%%%
\section{Numerical results and comparison with observations}\label{sec5}
%%%%%%%%%%%%%%%%%%%%%%%%%%%%%%%

\noindent In this section we show the results of the numerical calculations. We focus in particular on the maximal mass as a function of the parameter $\eta  Q_{p}^{2}$, defined by \eqref{qupi}, and of the EOS. As mentioned in the previous section, the limit $\eta Q_p^{2}=0$ does not lead to GR, however it gives very similar maximal mass values as in GR, so it can be used as a reference in the plots below. In general, we find that, for non-vanishing $\eta Q_p^{2}$, the maximal mass is generically lower than in GR when $\eta>0$ and larger when $\eta<0$.

\begin{figure}
  \includegraphics[scale=1]{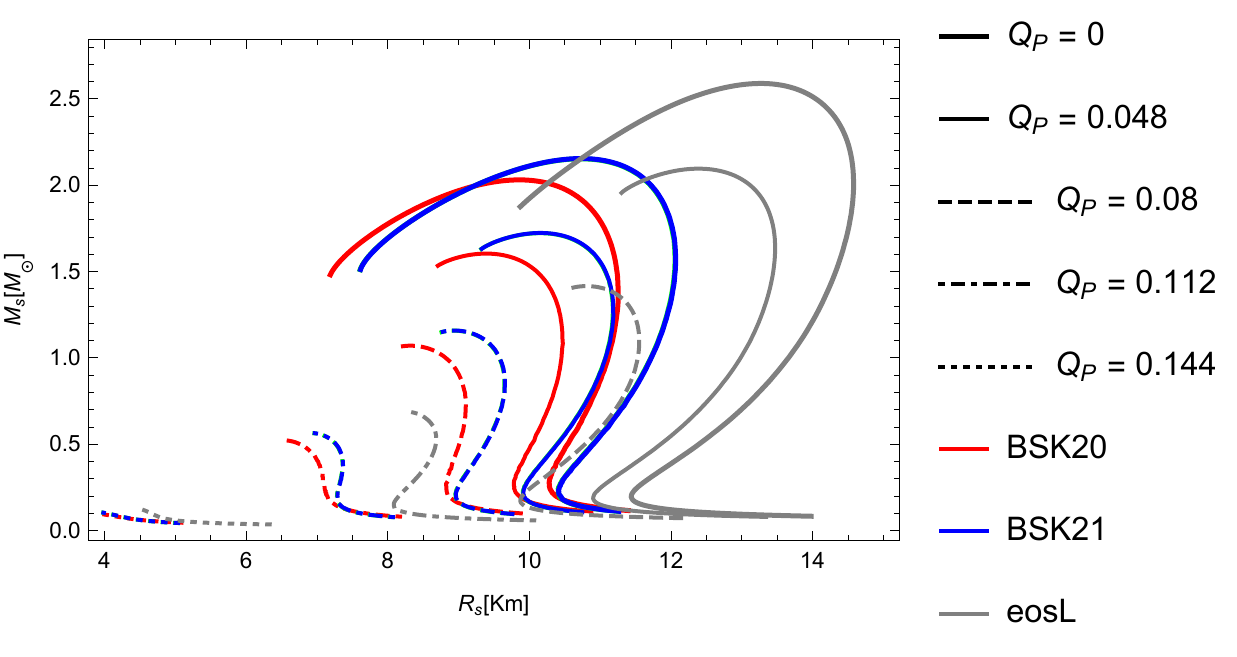}
  \includegraphics[scale=1]{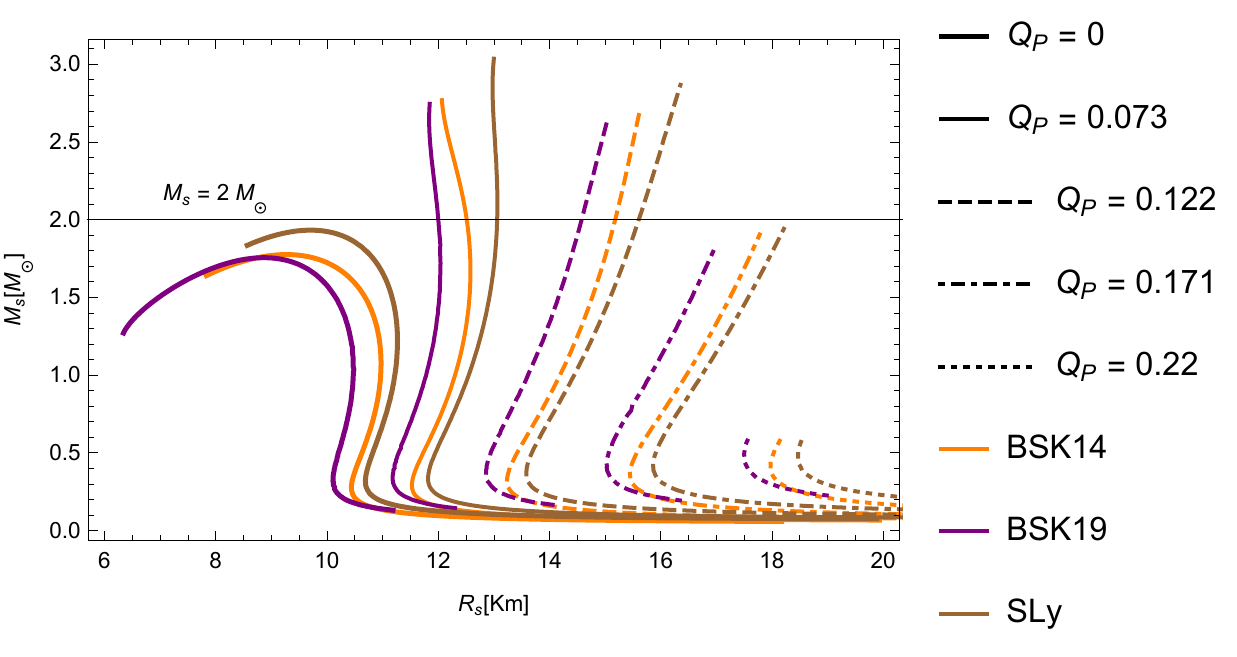}
 \caption{Mass-Radius relations for various EOS and values of $Q_p$. The upper (lower) panel shows the case $\eta>0$ ($\eta<0$). For $\eta<0$, we select the curves (hence the equations of state) that cannot reach a mass of  $2 M_\odot$ in GR.}
 \label{MRrel}
\end{figure}

In particular,  for $\eta>0$, we find that the mass-radius curve eventually terminates at the maximal value for the mass, which depends on the chosen EOS and on the central pressure. This is illustrated in Fig. \ref{MRrel}, where we show a few mass-radius curves for different values of $Q_p$ (we set $|\eta|=1$).

In the case $\eta<0$, there are no solutions when $Q^2|\eta| < 12\kappa P_{c}/(2\rho_{c}-3P_{c})$. When $Q^2|\eta| > 12\kappa P_{c}/(2\rho_{c}-3P_{c})$, we find that the the maximal mass increases according to $Q_p^2$, which can have arbitrarily large values. However, as we have seen above, negative values of $\eta$ seems to be unphysical from a quantum point of view, thus we do no really trust this result. 

In the case $\eta>0$, the compactness of the star as a function of the mass increases with the value of $Q_p$, but the soltions ceases to exist well before the black hole compactness is reached. For $\eta<0$, it is the other way around and the compactness decreases at fixed mass.

\begin{figure}
 \includegraphics[scale=.9]{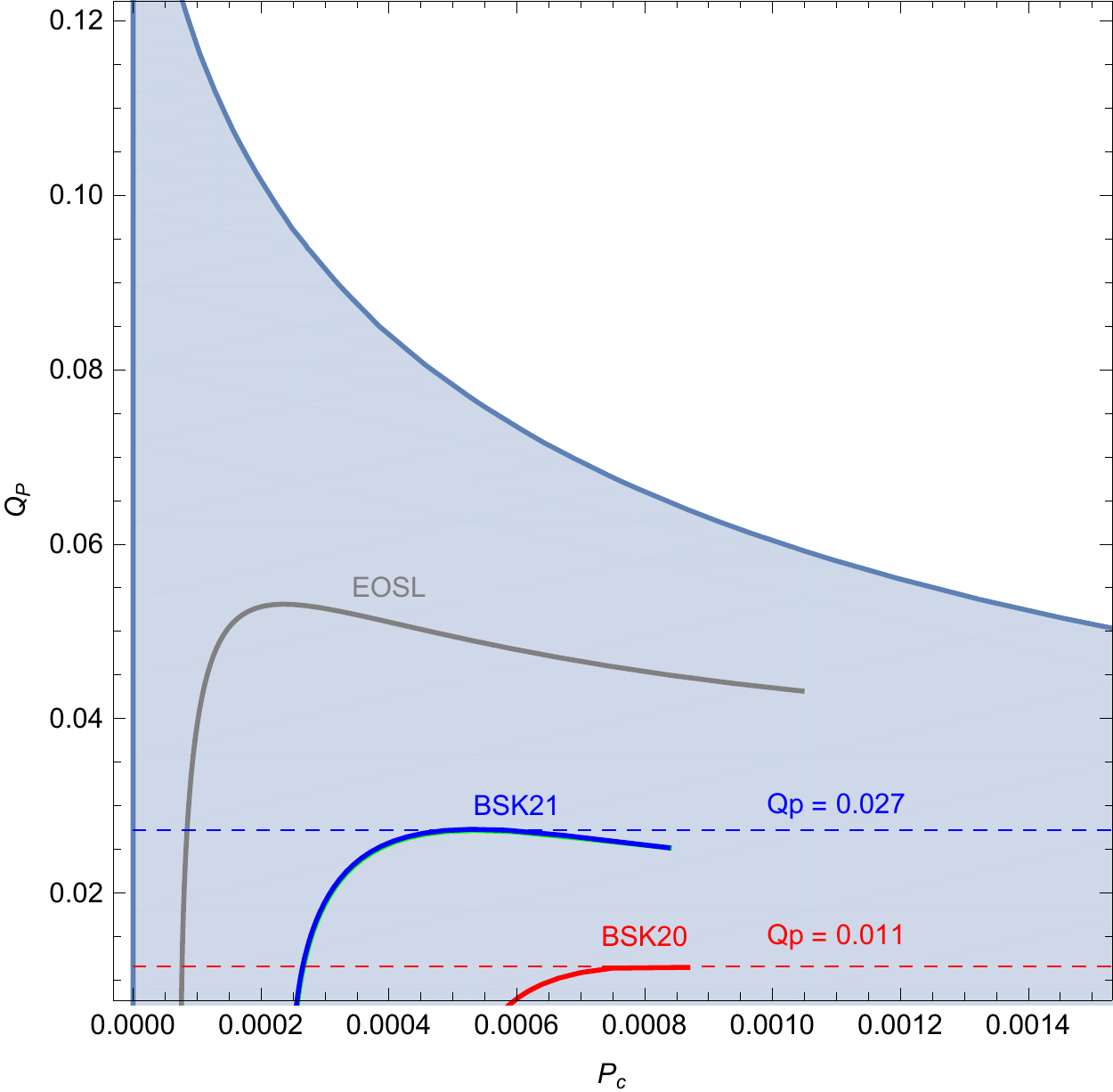}
 \caption{Constant mass curve with $M = 2M_\odot$ in the $P_c - Q_p$ plan. The shaded region is the region leading to compact solutions for all the considered EOS. The endpoints of the curve correspond to the maximal value of the central pressure available in the EOS tables that we use.}
 \label{M2Ms}
\end{figure}

We also note, as can be infered from figure \ref{M2Ms} that for $\eta>0$, the central pressure leading to a given value of the mass increases with the $Q_p$, while it is again the other way around for $\eta<0$. This is consistent with the result for the effective Planck mass \eqref{kappaeff} (recall that $G\propto 1/\kappa$) where it can be seen that $\eta>0$ increases the effective gravitational coupling, leading to a stronger gravity, with the consequence that matter is more compacted with increasing values of $Q_p$. In the case $\eta<0$, gravity becomes weaker and the pressure takes the lead and dilutes the configuration.

\begin{figure}
 \includegraphics[scale=1]{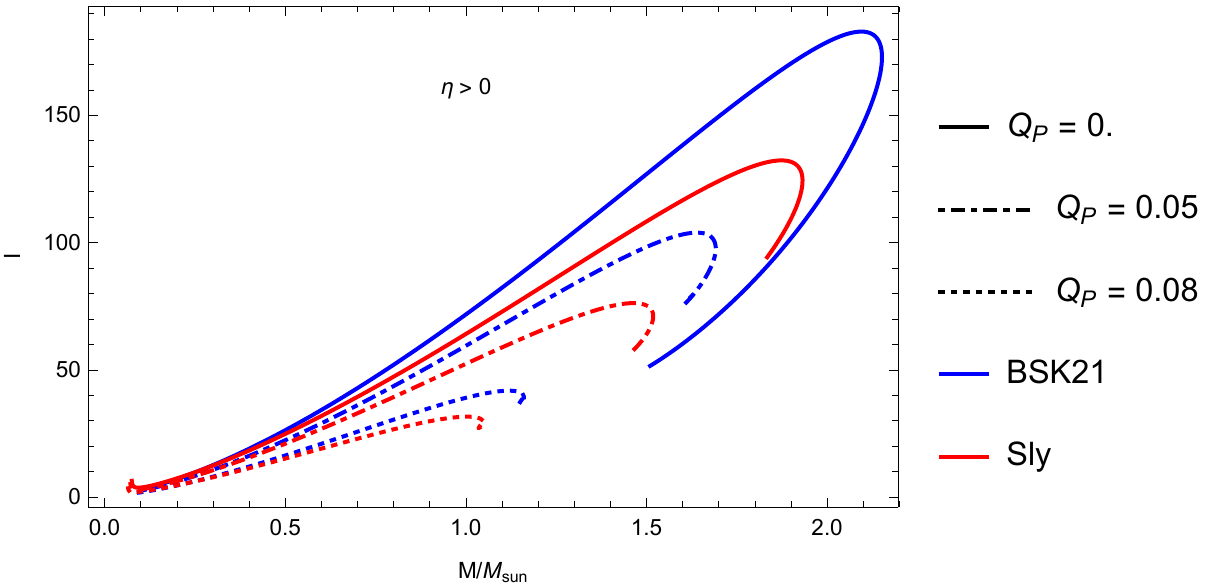} 
 \includegraphics[scale=1]{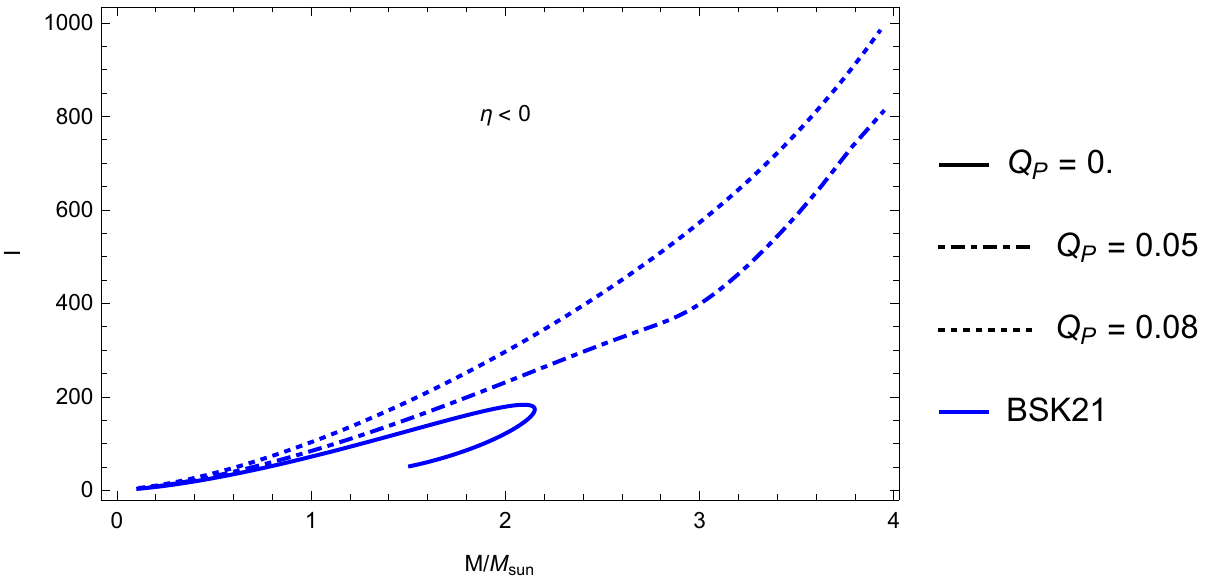}
 \caption{The mass - inertia curves for BSK20 and different values of $Q_p$ with $\eta >0$ (upper panel) and $\eta<0$ (lower panel).}
 \label{fig:inertia-mass} 
\end{figure}

In Fig.\  \ref{fig:inertia-mass}, we plot the inertia as a function of the mass, as a result of the equation \eqref{eq:inertia}. In the case $\eta >0$, the inertia is a strictly increasing function of the mass of the star up to a maximal value (depending on the value of the $Q_p$ parameter), where the curve turns-back. From this point on,  both inertia and mass decrease. Note also that the values of  the mass and of the inertia of the turning point  decrease while increasing the value of $Q_p$ . 

On the other hand, for the case $\eta<0$ the situation is a bit different. For small values of $Q_p$, the pattern of the curve is similar to the case $\eta>0$. But, for larger $Q_p$, the value of the inertia for a given mass \textit{increases} without turning point.  So the inertia becomes a monotonic growing function of the mass.

The most massive pulsar known to date is  PSR J0348+0432 with a mass of $2.01 \pm 0.04$ solar masses and with an orbital period of 2 hours and 27 minutes. In the case $\eta>0$, the existence of stars with this mass in our model  is not guaranteed for all values of the parameter $Q_p^2\eta$. We numerically found the constant mass curves in the plane $P_{c}-Q_{p}$ (namely central pressure vs $Q_{p}$) with $M=2M_\odot$. We find  that, depending on the specific EOS, these curves admit a maximum value of $Q_{p}$. This is illustrated in Fig. \ref{M2Ms}. For the specific cases of $BSk21$, $BSk20$ and $EOSL$ we observe that configurations with masses of the order of $M=2M_\odot$ can be obtained. This impose a constraint on the maximum value that $Q_p$ can take. For the remaining equations of state, namely, $BSK14$, $BSK19$ and $SLy$ the maximal mass cannot reach $M=2M_\odot$ for $\eta>0$ (for $\eta<1$ we have seen that the mass in basically unbounded for most EOS).

\subsection{Gamma Ray burst repeater redshift}
\begin{figure}
 \includegraphics[scale=.9]{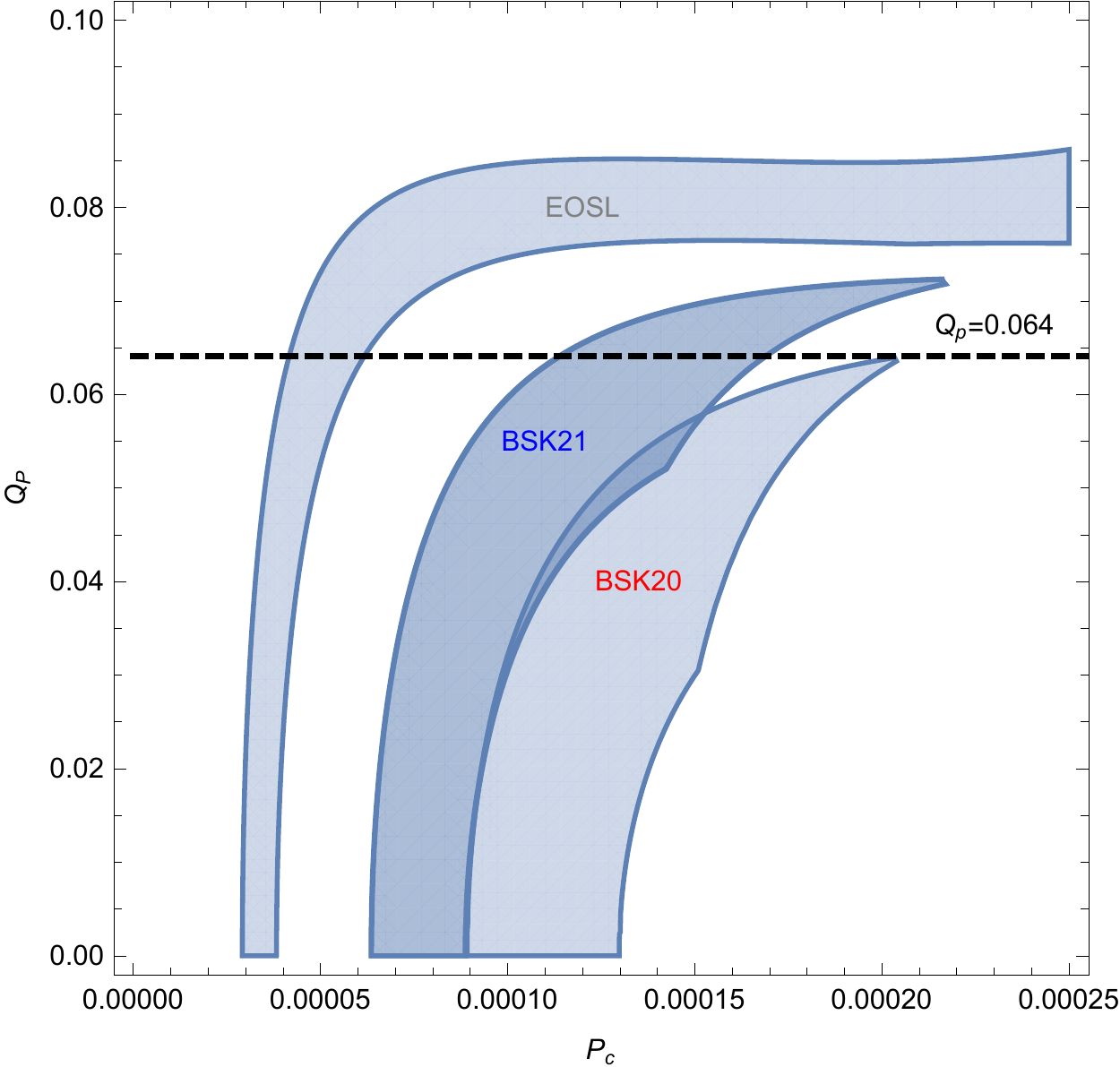}
 \caption{Redshift range for $z = 0.23 \pm 0.07$, $1.3<M<1.5 M_\odot$, and $\eta=+1$  in the $P_c - Q_p$ plane. The shaded regions correspond to the given $z$ and various EOS.}
 \label{redshift}
\end{figure}

\noindent Among the physical quantities of interest, there is the gravitational redshift $z$ of the photons emitted from the surface of a neutron star. For static configurations, it is defined by \cite{Haensel}
\be
z = \left( 1- \frac{2M}{R} \right)^{-\frac{1}{2}} - 1.
\ee
The value of the surface redshift  can be inferred from spectroscopic studies of gamma ray burst from the class of gamma-ray burst repeater. In our case, we  use the data of the gamma-ray burst GRB 790305 from the soft gamma-ray burst repeater SGR 0526-066 \cite{Higdon}. Interpreting the emission line at $430 \pm 30\ keV$ as the annihilation of electron-positron pairs, leads to 
\be
z= 0.23\pm 0.07.
\label{eq:surf-red}
\ee
In GR, the EOS $BSk19, 20, 21$ and $SLy$ are consistent with this value for neutron stars with mass around $1-1.5 M_\odot$. We computed the constant $z$-curves in the $P_c - Q_p$ plane, and we found that our model yields similar results for values of $Q_p$ constrained by the mass range only, see Fig.\ \ref{redshift}.

Note that we assume here that the line is emitted at the surface of the star.
Assuming instead that it originated around $10\%$ further away would change the
value of the redshift OF at most $10\%$, which is smaller than the error bars in
(59) that are of the order of $30\%$. In the following, the redshift is computed for static
configurations, which is unaffected by first order corrections due to rotation. 

We find that surface redshift observations lead to milder constraints on $Q_p^2\eta$ than the maximal mass. This is due to the fact the surface redshift is related to the mass-radius ratio, and the observed surface redshift range that we use is typical of neutron stars with mass lower than $2 M_{\odot}$. However, it is worth pointing out that, within the constraint provided by the observation of a 2 solar masses neutron star, the shift symmetric Horndeski model that we consider is still compatible with surface redshift observations, so there is no tension. More specifically, we find that neutron stars with masses in the conservative range of $1.3 - 1.5 M_\odot$ are compatible with surface redshift in the range $z = 0.23 \pm 0.07$, using equations of state that are not excluded in GR, \cite{Fantinia:2013}.

\begin{figure}
 \includegraphics[scale=.9]{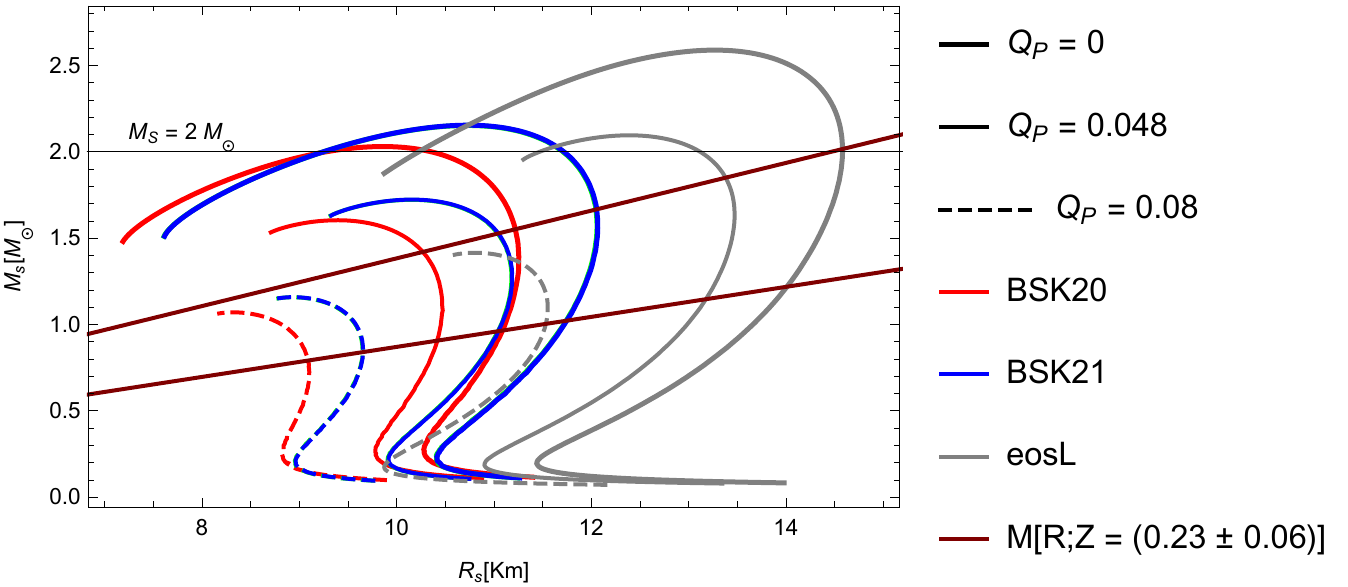} 
 \includegraphics[scale=.9]{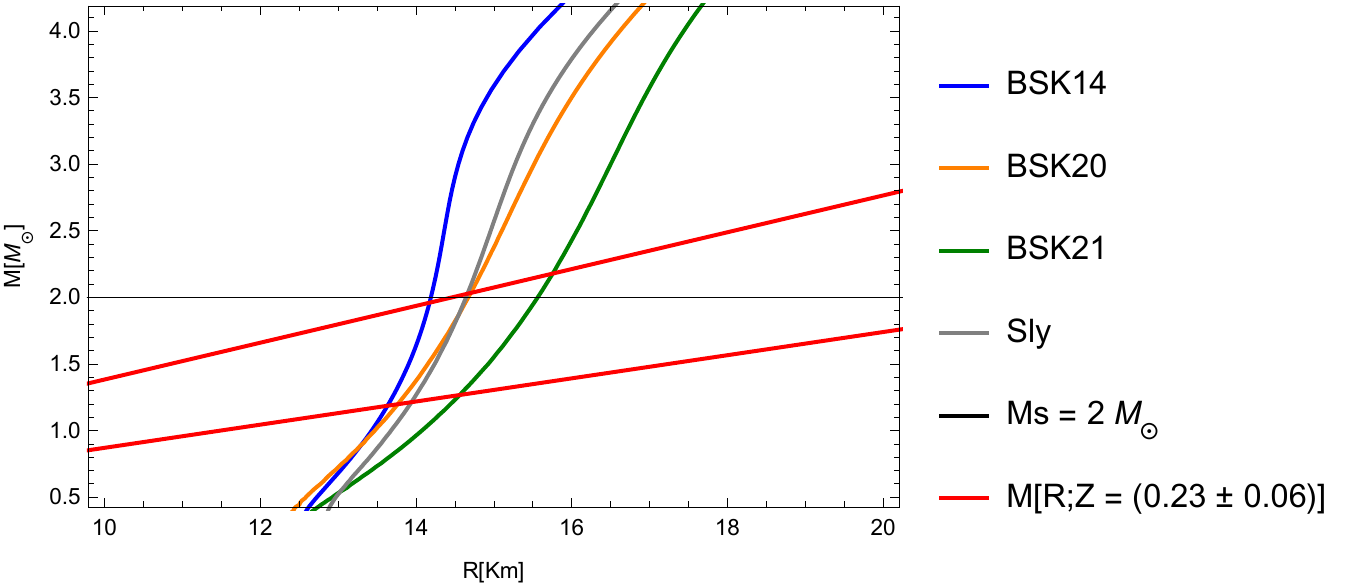}
 \caption{Mass-radius relations for $\eta>0$ (lower: $\eta<0$), together with the constant surface redshift curves for $z = 0.29$ and $z = 0.16$ for three different equations of states and different values of $Q_p$ (for $\eta<0$, $-Q_p\eta = 0.104$). With these choices of $Q_p$, the configurations leading to surface redshift in the measured range has the expected typical values. }
 \label{fig:mr-redshift}
\end{figure}

In Fig.\ \ref{fig:mr-redshift}, we superpose the mass-radius curves for a given EOS, with constant surface redshift  curves corresponding to the minimum and maximum observed $z$. This figure shows the expected mass range in our model with the surface redshift infered from GRB 790305.

\subsection{Estimation of the inertia of the Crab pulsar}

\noindent The moment of inertia of neutron stars is not yet measured, but lower bounds can be inferred from pulsar timing observations \cite{Bejger:Haensel, Haensel:2007}. An interesting example is the historical Crab pulsar. 
In general, according to GR, and with the assumption that the energy loss of the pulsar spin mainly goes into the acceleration of the nebula, leads to 
\be
I \geq \frac{M_{neb} v\dot v}{\Omega|\dot\Omega|}.
\ee
Here, $M_{neb}$ is the mass of the nebula, $v$ its expansion velocity, $\dot v$ its acceleration, $\Omega$ the angular frequency of the pulsar, and $\dot \Omega$ its time derivative.

For the the Crab nebula, it is estimated that $M_{neb} = 4.6 \pm 1.8 M_\odot$. Within GR, this corresponds to the following lower bounds for the inertia:
\bea
& M_{neb} = 2.8 M_\odot \Rightarrow I_{Crab} = 1.4 \times 10^{45} g\, cm^2,\non\\
& M_{neb} = 4.6 M_\odot \Rightarrow I_{Crab} = 2.2 \times 10^{45} g \,cm^2,\non\\
& M_{neb} = 6.4 M_\odot \Rightarrow I_{Crab} = 3.1 \times 10^{45} g \,cm^2.\non\\
\eea 
On the other hand, the Crab pulsar is expected to have a mass around $1.3-1.4 M_\odot$, from core collapse simulations in GR. This value of the mass is close to the Chandrasekhar limit (around $1.4 M_\odot$) for white dwarfs, above which an instability leading to the core collapse sets in. The collapsing core then forms a neutron star with a mass close to this value, up to matter ejected after bouncing on the forming star. The ejected mass from the collapsing core can be estimated in core collapse simulations. 
Although such simulations should be repeated in the precise model that we study for a consistent analysis, we don't expect the conclusion to change much, because potential deviations due to modification of gravity should affect only the strong field regime of the process at the end of the collapse. However, in order to remain as core-collapse model independent as possible, we do not impose a strict range for the mass of the pulsar, and use a (large) inertia range estimated from the mass of the Crab nebula remnant mass (as is explained in \cite{Fantinia:2013}). The range of values for the  inertia that we are dealing with reduces to
\be
I_{Crab} \in [1.4, 2.2] 10^{45} g/cm^2 = [104, 163.4],
\ee
where the last term gives the same values in units where $G=c=1$.

For our purpose of checking if the model we study is able to produce a configuration with a mass and inertia in the rough estimates given above, we will take these estimates for granted, and build the domain of existance of stars in these mass and inertia range.

\begin{figure}
  \includegraphics[scale=.7]{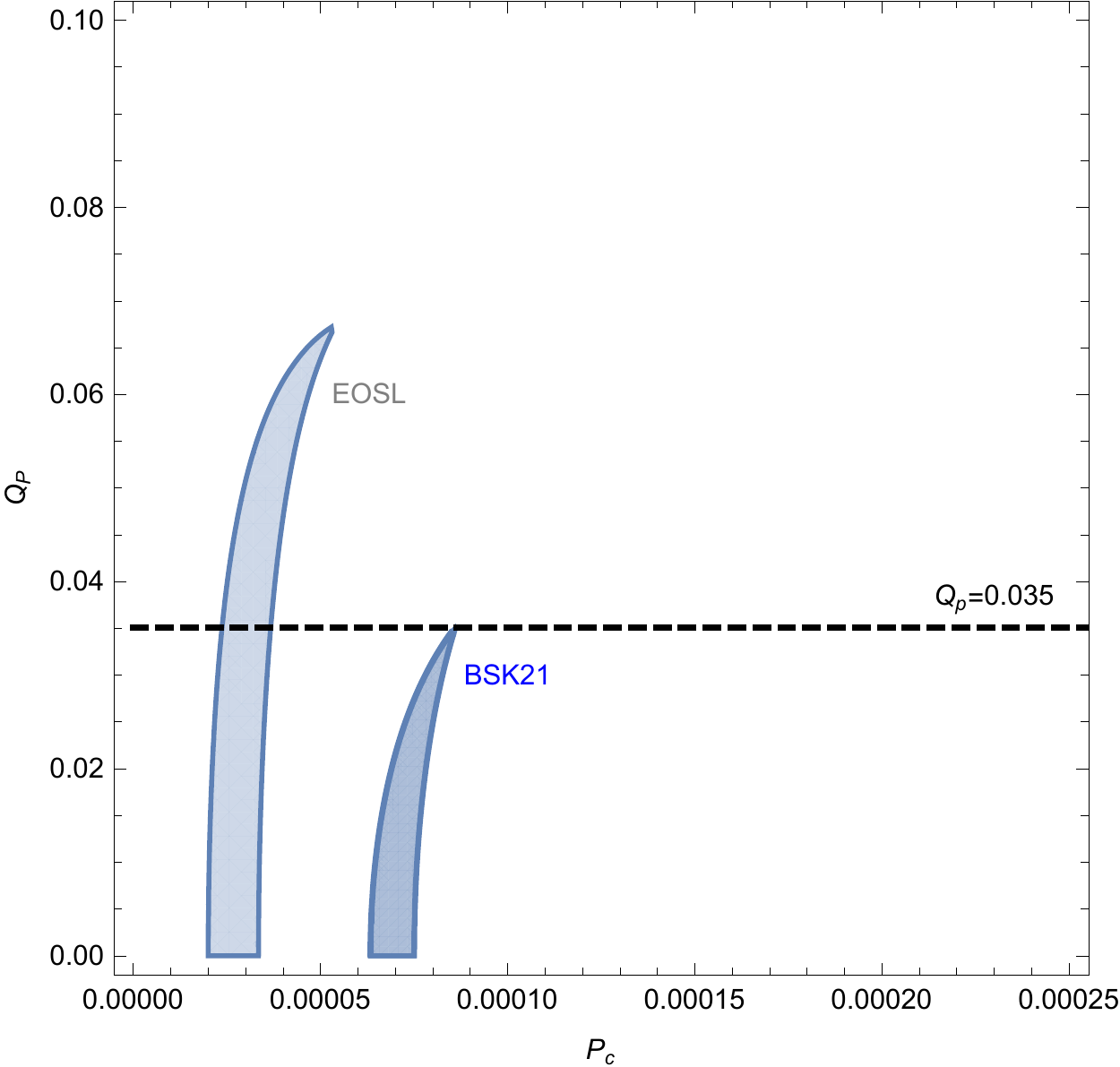}
  \includegraphics[scale=.7]{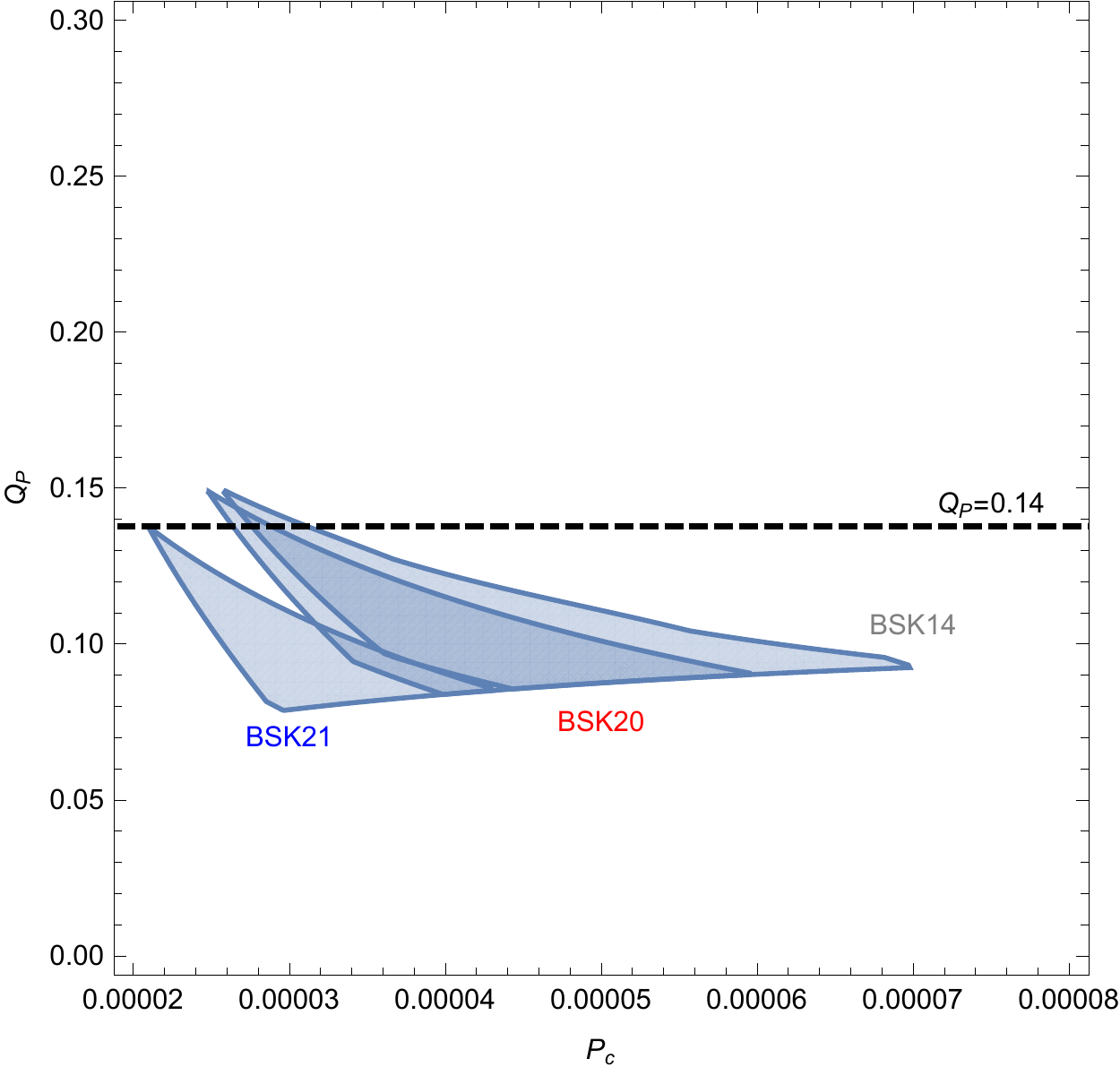}
  \caption{Plot of the region where the mass of the Crab pulsar is in the range $[1.3,1.5] M_\odot$ and the inertia is in $[1.4, 2.2] 10^{45} g/cm^2$ for $\eta>0$ (upper panel) and $\eta<0$ (lower panel) for some EOS, in the $P_c - Q_p$ plane. The horizontal line corresponds to our best constraint from the observation of the neutron star of two solar masses. The left boundaries of the existence domain is the constant inertia curve ( smaller bound for $\eta>0$ and larger for $\eta <0$) while the right boundary is a constant mass curve (the largest bound for $\eta>0$ and the smallest for $\eta<0$.  }
  \label{fig:inertia-region}
 \end{figure}

In figure \ref{fig:inertia-region}, we show the region in the $Q_p - P_c$ plane, where the inertia is in the range described above, and the mass is in the range $M = 1.3 - 1.5 M_\odot$. As one can see, this region is compatible with the strongest constraint that we found above for $Q_p$. This is the case for all the EOS that we considered. Interestingly, the largest value of $Q_p$ in the region plot is close to our best constraint, suggesting that inertia measurement might constraint the model parameter almost as good as maximal mass measurements. Note that for $\eta>0$, the boundary of the domain are given by the constant inertia curve with the smallest value (left boundary) and the constant mass curve with largest mass (right boundary). Accepting $1.4 M_{\odot}$ as an upper mass limit for the Crab pulsar and the lower bound of the inertia as a lower limit leads to the result presented here, even with larger mass and inertia ranges. 
Note that for the case $\eta<0$, the argument is weaker because the domain is bounded by the minimal mass and maximal inertia.

%%%%%%%%%%%%%%%%%%%%%
\section{Conclusion}\label{sec6}
%%%%%%%%%%%%%%%%%%%%%

\noindent We studied slowly rotating neutron stars in the shift-symmetric sector of Horndeski gravity with realistic equations of state, modelling dense nuclear matter. The model that we built describes well most of the observed pulsars, with the exception of millisecond pulsar that are in the rapid rotation regime. 

We also investigated the cosmological solutions in the  the same theory. We found that cosmological and astrophysical configurations  are consistent  provided the usual kinetic term parameter ($\alpha$) is vanishing together with the cosmological constant. This choice leaves  only one free parameter, namely the derivative coupling strength $\eta$. The scalar field is chosen to  be linear in time, providing an additional degree of freedom $Q_p$ that effectively combines with $\eta$, leading to a single model parameter (apart from Newton's constant) $Q_p^2 \eta$.

We derived constraints on $Q_p^2 \eta$ by requiring that our model reproduces the mass of the largest neutron star observed so far and checked the consistency with other constraints. For example, we showed that there exist neutron stars with a surface redshift compatible with typical measurements. Additionally, we found that the inertia of our neutron star is compatible with the inertia estimates for the Crab pulsar, with masses in the expected range from core collapse scenario. 

We confirmed previous suspicions that the external structure of the spacetime is unaltered by the scalar field, in the slow rotation limit, leading to a nontrivial effect of the gravity modification inside the star only. From this point of view, the nonminimal kinetic coupling model considered here shares some similarities with models having a modified matter coupling, as is the case of Eddington-inspired Born Infeld (EiBI) gravity \cite{Banados:2010ix, Pani:2011mg, Delsate:2012ky, Pani:2012qb}. However, the Horndeski alternative is more promising since its equations structure is different and should not lead to surfaces singularity as in the EiBI model \cite{Pani:2012qd}.  

Since the configurations that we studied admit exactly the same exterior solutions as in GR, binary pulsar tests are expected to be valid with the shift-symmetric sector of Horndeski gravity. Of course, regarding the gravitational wave emission, we expect a modification due to the scalar field and the non-minimal coupling, but the geodesic motion itself, as long as backreaction effects are expected to be negligible. In conclusion, for slowly rotating solutions, the effect of the non-minimal kinetic coupling is to effectively modify the internal structure of the star. This is why we expect tests based on binary pulsar observation to succeed in this model. In order to fully address this question, we plan to study spherically symmetric perturbations of compact stars in the shift symmetric Horndeski model with non-minimal kinetic coupling elsewhere. 

Finally we comment on the constraints derived in this paper. In the case $\eta>0$, the constraint is provided by the EOS BSK21 and it is given by $Q_p^{2}\eta \leq 0.027$. Note that the EOS BSK20 leads to a more stringent constraint ($Q_p^2\eta \leq 0.011$). However, as in GR, the maximal mass for BSK20 is only slightly above 2 solar mass, and this is the reason why we tend to not consider this constraint as the most conservative one.

\section{Acknowledgement}
\noindent A.C would like to thank interesting discussions with C.\ Charmousis and E.\  Babichev during his visit to Orsay. T.D. and L.D. acknowledges useful discussions with E.\ Barausse. M.R. acknowledge useful discussions with S. Zerbini and L. Vanzo. We are grateful to E.\ Berti for interesting discussions. A.C. work is supported by FONDECYT project N\textordmasculine3150157. T.D. gratefully acknowledges the Belgian FRS-FNRS for financial support, as well as the Wallonie-Bruxelles International Grant of Excellence for partial financial support.

% \bibliography{horsr}

\end{document}